\documentstyle[aps,prl,amsmath,amssymb,epsfig,twocolumn]{revtex}
\newcommand{\ph}{\varphi}
\begin{document}
\draft
\title{$\pi-0$ Transition in Superconductor-Ferromagnetic-Superconductor
Junctions}
\author{N.\,M.\,Chtchelkatchev$^{+}$,
W.\, Belzig$^\dag$, Yu.\,V.\,Nazarov$^*$, and C.\,Bruder$^\dag$}
\address{$^+$L.~D. Landau Institute for Theoretical Physics RAS,
117940 Moscow, Russia\\
$^\dag$ Departement Physik und Astronomie, Universit\"at Basel,
Klingelbergstr. 82, 4056 Basel, Switzerland\\
$^*$ Department of Applied Physics and Delft Institute of Microelectronics and Submicrontechnology,\\ Delft
University of Technology, Lorentzweg 1, 2628 CJ Delft, The Netherlands}
\date{\today}
\maketitle
\begin{abstract}
\bf Superconductor-Ferromagnetic-Superconductor (SFS) Josephson junctions are known to exhibit a transition between
$\pi$ and $0$ states. In this note we find the $\pi-0$ phase diagram of an SFS junction depending on the
transparency of an intermediate insulating layer (I). We show that in general, the Josephson critical current is
nonzero at the $\pi-0$ transition temperature. Contributions to the current from the two spin channels nearly
compensate each other and the first harmonic of the Josephson current as a function of phase difference is
suppressed. However, higher harmonics give a nonzero contribution to the supercurrent.
\end{abstract}

\vspace{-0.4truecm} \narrowtext

In the last years, many interesting phenomena were investigated in Superconductor (S) - Ferromagnetic (F) -
Superconductor Josephson contacts. One of the most striking effects is the so called $\pi$-state of SFS junctions
\cite{pi_1,pi_2,Buzdin,Ryazanov} in which the equilibrium ground state is characterized by an intrinsic phase
difference of $\pi$ between the two superconductors. Investigations of $\pi$ junctions have not only academic
interest, e.g., in \cite{feigel_1,feigel_2} a solid-state implementation of a quantum bit was proposed based on a
superconducting loop with $0$ and $\pi$ Josephson junctions.

The existence of the $\pi$-state in an SFS junction was recently experimentally demonstrated by the group of
Ryazanov \cite{Ryazanov}. In this experiment, the temperature dependence of the critical current was measured. At a
certain temperature the critical current was found to drop almost to zero, this has been interpreted as the
transition from the $0$ to the $\pi$ state. The transition temperature $T_{\pi 0}$ was shown to exhibit a strong
dependence on the concentration of ferromagnetic impurities, i.e., on the exchange field $E_{\rm ex}$ in the
ferromagnetic film.

In this note, we present a theory of the $\pi-0$ transition in short SFS junctions. Our goal is to understand which
parameters (exchange field, temperature,...) stabilize a $\pi$ state and how the phase diagram looks like. We
investigate the current-phase relation and the critical current near the transition to the $\pi$-state. Most
importantly, we find that in general, the critical current is not zero at $T_{\pi 0}$, and it may not even reach a
local minimum. The identification of the critical current drop and the $\pi-0$ transition is only possible if the
current is given by the standard Josephson expression $I(\ph)\propto \sin(\ph)$, which is valid for the limiting
case of tunnel barriers only. Even if the main contribution to the current is of this form, the higher harmonics
contribution $I(\ph)\propto \sin(2\ph)$ would not vanish at $T_{\pi 0}$. Consequently, $I_c\neq 0$ at the transition
point.

We consider SFS junctions in the ``short'' limit defined by $\hbar/\tau\gg\Delta(T=0)$; here, $\tau$ is the
characteristic time needed for an electron to propagate between the superconductors. In this case, we can employ a
powerful scattering formalism \cite{been1} which allows one to express the energies of Andreev states in the
junction in terms of the transmission amplitudes of the junction in the normal state. These Andreev states give the
main contribution to the phase-dependent energy of the junction, therefore $I(\ph)$ can be calculated. Any junction
is characterized by a set of ``transport channels'' labelled by $n=0,1,\ldots, N$, each channel is characterized by
the transmission coefficient $D_n$. If one disregards ferromagnetism, the Andreev levels are degenerate with respect
to the spin index $\sigma$. Their energies are given by $E_{n\sigma}=\pm\Delta(1-D_n\sin^2(\ph/2))^{1/2}$.

\begin{figure}[htb]
\begin{center}
\epsfxsize=60mm \epsffile{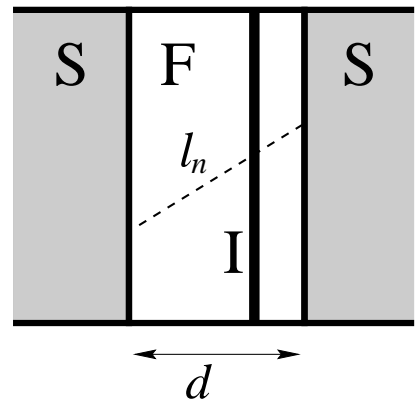}
\end{center}
\caption{Junction configuration. The current flows from one superconductor (S) to the other through the
ferromagnetic (F) layer (width $d$) which a scattering region denoted by I. The exchange field is supposed to be
parallel to the SF boundaries.} \label{fig1}
\end{figure}

We generalize the scattering approach to cover SFS junctions. In this case, the phase of the transmission amplitudes
also becomes important. To see this, we introduce the parameter $\gamma$: $\cos(\gamma(\ph)) \equiv
1-2D_n\sin^2(\ph/2)$. If we assume that ferromagnetism does not change the transport channels, the energies of the
Andreev states become
\begin{gather}
\label{A_levels_1} E_{n\sigma}(\ph)=\Delta\left|\cos\left(\frac{\gamma(\ph)
+(\Phi_{n,\sigma}-\Phi_{n,-\sigma})}2\right)\right|\; ,
\end{gather}
where $\Phi_{n,\sigma}$ is the phase of the transmission amplitude for an electron with spin $\sigma$ in the channel
$n$. Thus we observe that the different phase shifts for different spin directions result in a spin-dependent {\it
energy} shift of the Andreev states. To specify the model further, we consider the layout shown in Fig.~\ref{fig1}.
It consists of two bulk superconductors, a ferromagnetic layer with exchange energy $E_{\rm ex}\ll E_F$, and a
scattering region denoted by I. We assume that the F-layer is ballistic and that the order parameter $\Delta$ is
constant in the superconductors: $\Delta(x)=\Delta e^{\pm \ph/2}$, and $\Delta(x)=0$ in F.

We believe that the model considered is quite a general one. It is applicable to quasiballistic SFS multilayers
(recently a quasiballistic SF junction was prepared by Kontos {\it et al.} \cite{Kontos}) with either specular or
disordered interfaces \cite{Nazarov}; also to Josephson junctions where electrons tunnel through small ferromagnetic
nanoparticles \cite{Gueron,fogelstrom}. We shall restrict ourself to the case, when the width of the scattering
region is much smaller than the width $d$ of the junction. The transport channels can be associated with different
incident angles. Then $\Phi_{n,\sigma}-\Phi_{n,-\sigma}=\sigma\pi(2E_{\rm ex}d/\pi\hbar v_F)l_n=\sigma\pi\Theta
l_n$, where $l_n>1$ is the length of a quasiparticle path between the superconductors divided by $d$, see
Fig.~\ref{fig1}. Formula \eqref{A_levels_1} reproduces the energy spectrum obtained in the limiting case $D=1$ in
\cite{Falko}.

The contribution to the free energy of the junction which depends on $\ph$ is given by
\begin{gather}
\label{Omega} \Omega(\ph)=-T\sum_{n,\sigma}\ln \left [\cosh\left(\frac{E_{n\sigma}(\ph)}{2T}\right )\right]\, .
\end{gather}
The continuous spectrum is neglected in \eqref{Omega}; one can easily check that it gives a $\ph$-independent
contribution to the free energy. The summation over the channels $n$ can be evaluated by converting the sum to an
integral: $\sum_n\ldots=\int dl\rho(l)$, where $\rho(l)=\sum_n\delta(l-l_n)$ and $\int \rho(l)dl=N$, the number of
channels. If there is only one channel in the junction, the weight function $\rho$ defined above reduces to
$\delta(l-1)$. If, on the other hand, the number of channels $N$ is much bigger than unity,
$\rho(l)=2N/l^3\theta(l-1)$. (We assumed $D$ to be independent on $n$.) A similar distribution of $l$ can be found for
SFS junctions with disordered boundaries, see \cite{Nazarov}. At some points of these notes, we shall use the
distribution $\rho(l)= N\delta(l-1)$, since it allows us proceed analytically, and the results obtained with it are
qualitatively the same as with the other distributions.  We shall refer to this distribution as the
$\delta$-distribution. (When $\rho(l)= N\delta(l-1)$,  our parameter $\Theta$ is closely related to spin-mixing angle
introduced in \cite{fogelstrom}.)

Which exchange field in F is sufficient to ensure that the SFS junction can be put into a $\pi$-phase by changing
the temperature? The $\pi$-state is the result of the ferromagnetic exchange field in the F-layer. If it is too
small, then the junction will remain in the $0$-phase at all temperatures. We show below which values of exchange
field and temperature guarantee that the junction will be in the $\pi$-phase.

\begin{figure}
\epsfxsize=80mm \epsffile{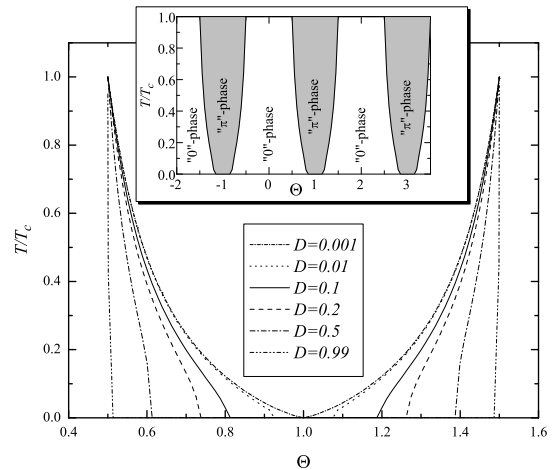} \caption{The temperature of the $\pi-0$ transition at zero current versus the
dimensionless exchange field $\Theta=2E_{\rm ex}d/\pi\hbar v_F$ at different values of junction transparencies $D$.
Only trajectories with $l=1$ are taken into account which is justified if either one channel or many channels are
present. Inset: phase diagram of the junction at $D=0.1$. The gray regions correspond to the $\pi$-phase, the white
regions to the $0$-phase.} \label{fig2}
\end{figure}

In an equilibrium situation with zero current, the temperature $T_{\pi 0}$ separating the $\pi$ and $0$ phases is
determined from the condition that the free energy $\Omega$ reaches its minimum at $\ph=2\pi n$ and at $\ph=\pi+2\pi
n$, $n=0,\pm 1,\ldots$ (the free energy of an ordinary junction has a global minimum at $\ph=2\pi n$). The numerical
solution of this equation for $T_{\pi 0}(D)$ is shown in Fig.~\ref{fig2}. Here and below, we use the approximation
$\Delta(T)/\Delta(0)=\tanh(1.74\sqrt{T_c/T-1})$ in doing numerical calculations.
If $D=1$, the $\pi$-phase can exist only in the domain $2n+1/2<\Theta<3/2+2n$, $n=0,\pm 1,\ldots$. At finite $D$,
there are regions of $\Theta$ in which either the $\pi$-phase or the $0$-phase is stable for all temperatures,
$I_c(T)$ has no cusps in these regions. For $\Theta\to 1/2+n$, $T_{\pi 0}\to T_c$ for arbitrary transparency.

There are regions in the phase diagram where $\Omega$ has two minima, $\pi=2\pi n$ and $\ph=\pi(2n+1),n=0,\pm
1,\ldots$. We shall consider these regions below.

The $(T,\Theta)$ phase diagram of the junction is depicted in the inset of the Fig.~\ref{fig2}. The diagram is
periodic in $\Theta$ with period $2\pi$. It follows from the graph that large value of the exchange field
$\Theta=2E_{\rm ex}d/\pi\hbar v_F$ do not guarantee that the SFS junction is a $\pi$-junction.

Evidence of the existence of a $\pi$ phase in SFS junctions was experimentally demonstrated by the group of Ryazanov
\cite{Ryazanov}. The experimental curves $I_c(T)$ showed cusps at a certain temperature (which we shall denote by
$T_{\pi 0}$); the critical current at the cusp was close to zero. There are qualitative arguments in \cite{Ryazanov}
that the cusp corresponds to transition of the junction to the $\pi$ state and $I_c\equiv 0$ at the cusp. We agree
with the first statement, but disagree with the second. In our opinion, there is no qualitative argument for the
critical current to be zero at the temperature of the cusp. Our model gives qualitatively similar curves $I_c(T)$ to
those presented by Ryazanov {\it et al.}, but $I_c\neq 0$ at the cusp, where the junction undergoes transition
between 0 and $\pi$ states. This will be discussed below in more detail.

The Josephson current $I$ carried by the Andreev levels \eqref{A_levels_1} can be found from the free energy
\eqref{Omega} using the relation $I(\ph)=\frac{2e}{\hbar}\partial_\ph \Omega(\ph)$. Using
$\partial_\ph\gamma=D\sin(\ph)/\sin(\gamma)$, we obtain
\begin{gather}
\label{I_general} I(\ph)=\sum_\sigma\sum_n\frac{2e}{\hbar}\frac{D\sin(\ph)}{\sin(\gamma)}
\Delta\sin\left(\frac{\gamma+\sigma\pi\Theta l_n}2\right)\times\\\nonumber \tanh\left(\frac{\Delta}{2T}\cos
\left(\frac{\gamma+ \sigma\pi\Theta l_n}2\right)\right)\; .
\end{gather}
If $\Theta=0$ and $D=1$, Eq.~(\ref{I_general}) reduces to the usual formula for the Josephson current in a short
ballistic SNS junction, leading to the critical current $I_c=Ne\Delta/\hbar$ at $T=0$ \cite{been1,been2}.

If $D\ll 1$, we can proceed analytically in the calculation of $I_c(T)$. Then $\gamma\approx
2\sqrt{D}|\sin(\ph/2)|\ll 1$ and we can expand the Josephson current \eqref{I_general} in $\gamma$. This leads to
\begin{equation}
\label{I_D_small} I(\ph)=\frac{e\Delta D}{2\hbar}\sin(\ph)g(\Theta,T)\; ,
\end{equation}
where
\begin{eqnarray}
g(\Theta,T)&=&\int dl \rho(l)\left\{\cos\left(\pi\Theta l/2\right) \tanh \left( \frac{\cos\left(\pi\Theta
l/2\right)\Delta}{2T}\right) \right.\nonumber \\
&&- \left.\frac{\Delta \sin^2(\pi\Theta l/ 2)}{2T \cosh^2 \left(\cos\left(\pi\Theta
l/2\right)\Delta/2T\right)}\right\}\; . \label{g}
\end{eqnarray}
If $g>0$ in \eqref{g}, the junction is in the ordinary state. In the opposite case, the junction is in the $\pi$
state.

Solving $g(\Theta,T)=0$ gives us the transition temperature $T_{\pi 0}$. For $\rho(l)\propto \delta(l-1)$,
$g(\Theta,T)=0$ can be solved only in the domain $2n+3/2\leq\Theta\leq 1/2+2 n$, $n=0,\pm 1,\ldots$. As a consequence,
the $\pi$-phase exists only in these domains. If $\Theta\to 1/2+n$ then $T\to T_c$; $\Theta\to 1+2n$ leads to $T\to 0$.

In the region $|T-T_{\pi 0}|\sim D T_c$, the current is no longer given by Eq.~\eqref{I_D_small}. Higher harmonics in
$\ph$ have to be taken into account. Since the $n$-th harmonic is proportional to $D^n\sin(n\ph)$ (follows from
\eqref{I_general}) and $D\ll 1$, the second harmonic gives the main contribution to the Josephson current:
$I(\ph)\propto D^2\sin(2\ph)$. That means the critical current is not zero at the temperature of $\pi-0$ transition,
$I_c\propto D^2$. Near $T_{\pi 0}$ the currents of the spin channels $\sigma=\pm 1$ flow in opposite directions and
nearly compensate each other; therefore $I_c$ is suppressed.

Near the critical temperature $T_c$, we also can proceed analytically. Then we obtain from \eqref{I_general} for the
Josephson current:
\begin{gather}
\label{I_Tc} I(\ph)=\frac{e\Delta^2}{2T\hbar}\sin(\ph) \int dl\rho(l) D\sin(\pi\Theta l)\; .
\end{gather}
Using the $\delta$-distribution of the trajectory lengths, we find that the junction is in the $\pi$-state when
$1+2n<\Theta<2+2n,\,n=0,\pm 1,\ldots$.

\begin{figure}[htb]
\epsfxsize=80mm \epsffile{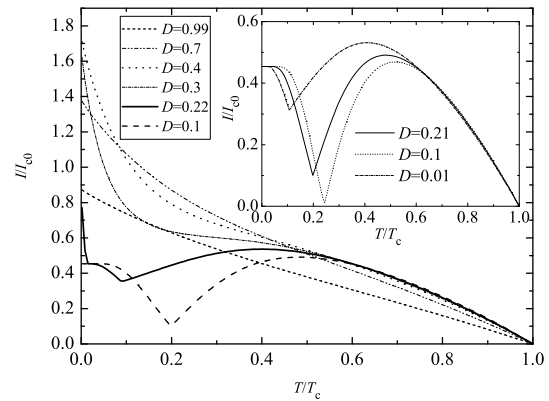} \caption{Dependence of the critical current on temperature for $\Theta=0.7$ at
different transparencies $D$. Cusps in the $I_c(T)$ curves indicate the $\pi-0$ transition. $I_c$ is always nonzero at
the cusp, see the inset. The plot of the critical current for $D=0.22$ has two cusps. At low temperature and $I\lesssim
I_c$, the junction with $D=0.22$ is in the $\pi$-state, at intermediate temperatures (between two cusps) -- in the
$0$-state, and for temperatures near $T_c$ -- again in $\pi$-state.} \label{fig3}
\end{figure}

Figure \ref{fig3} shows the typical dependence of the critical current on temperature.
The critical current in the figure is normalized to the critical current $I_{c0}=N(e\Delta/\hbar)(1-\sqrt{1-D})$ at
zero temperature and zero exchange field. The plot corresponding to $D=0.22$ exhibits two cusps. When the
temperature is low the junction is in the $\pi$-state; for intermediate temperatures (between the two cusps) the
junction will be in the $0$-state, for $T\lesssim T_c$ -- again in the $\pi$-state. (There is a schematic plot in
\cite{pi_1} where $I_c(T)$ of a SFS junction has many cusps with nonzero critical current in the cusps. However,
Ref. \cite{pi_1} does not provide an explanation of this fact).
The critical current for $D=0.3,0.4,0.7$ in Fig.\ref{fig3} is greater than the critical current $I_{c0}$
corresponding to zero temperature and zero $\Theta$. The exchange field enhances the Josephson current in the SFS
junction \cite{efetov}.

Below we shall investigate the $\pi-0$ transition when a $d.c.$ current $I<I_c$ is injected into the junction. We
shall pay attention to the regime when $I$ is smaller than $I_c$ at the cusp. Suppose that the temperature is
changed at fixed $I$. Then at the $\pi-0$ transition temperature, the phase across the junction will jump
approximately by $\pi$.

\begin{figure}
\epsfxsize=80mm \epsffile{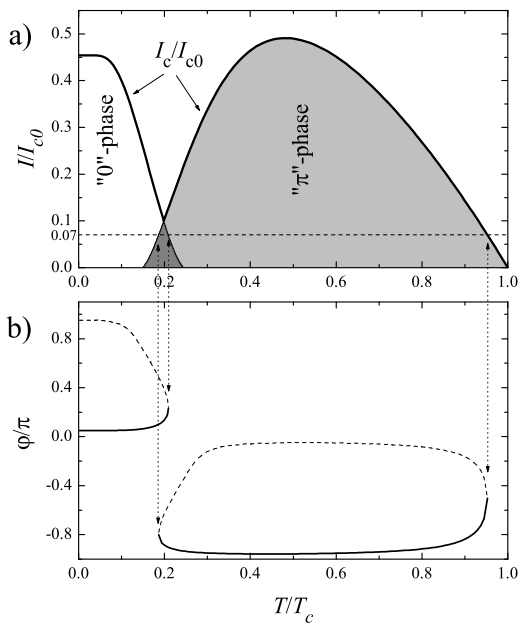} \caption{(a) The phase diagram of the junction for $D = 0.1$, $\Theta=0.7$. The light
gray region corresponds to the $\pi$-phase, the white to the $0$-phase. The Gibbs potential has two minima of $\ph$ in
$[-\pi,\pi]$ in the dark gray region. The critical current (solid thick line) is the upper boundary of the phase
diagram. (b) Temperature dependence of the phases corresponding to the d.c. current $I= 0.07I_{c0}$ (dashed line
parallel to the temperature axes in (a)). The thick solid line represents the stable solution of the equation
$I=I(\ph)$ (minimum of the Gibbs energy), the dashed curve exhibits the unstable solution (maximum of the Gibbs
energy). The equilibrium transition temperature ($I=0$) corresponds to $T/T_c=0.2$. } \label{fig4}
\end{figure}

There are several solutions $\ph(T)$ of the equation $I=I(\ph)$, where $I(\ph)$ is given by \eqref{I_general}. We will
assume that the damping is large, such that the phase is stabilized at one of the minima of the Gibbs energy
$\Xi(I,T,\ph)=\Omega(T,\ph)-\ph I\hbar/2e$ \cite{Licharev}. The phase values corresponding to the minima of the Gibbs
energy are depicted by solid lines in Fig.~\ref{fig4}b. If the temperature is increased from $T=0$, the phase will
continuously change with $T$ until $T$ reaches the dark gray region in Fig\ref{fig4}a where $\Xi$ has two minima of
$\ph$ in $[-\pi,\pi]$. Here the phase will choose one of the minima depending on the dynamics of the junction, which
depends on the properties of the external circuit. Outside of this region at higher temperatures, the phase will also
follow continuously adiabatic changes of $T$.

In conclusion, we investigated the phase transition between the $\pi$ and $0$ phases in a ballistic SFS junction
with a scatterer in the F layer. We calculated the $(T,\Theta)$ and $(I,T)$ phase diagrams of the junction. It was
shown that there is no reason for the critical current to be zero at the transition temperature $T_{\pi 0}$. The
currents of the two spin channels nearly compensate each other at $T_{\pi 0}$, and the current scales as
$D^2\sin(2\ph)$, $D\ll 1$ instead of $D\sin(\ph)$, as it does far from the transition.

We would like to thank V.\,V.~Ryazanov, G.\,B.~Lesovik, and M.\,V.~Feigel'man for stimulating discussions and useful
comments on the manuscript. Our research was supported by the Swiss National Foundation; N.~M.~C. is also supported
by the RFBR, project No. 01-02-06230, by Forschungszentrum J\"ulich (Landau Scholarship), and by the Netherlands
Organization for Scientific Research (NWO).

\end{document}